\documentclass[12pt]{article}

\usepackage{amsmath}
\usepackage{soul,color}
\usepackage{epsfig}
\usepackage{natbib}
\usepackage{makeidx}         % allows index generation
\usepackage{graphicx}        % standard LaTeX graphics tool when including figure files
\usepackage{multicol}        % used for the two-column index
\usepackage[bottom]{footmisc}% places footnotes at page bottom

\usepackage{array}
\usepackage{multirow} 
\usepackage{rotating}
\usepackage{multicol}
\usepackage{bm}
\usepackage{epsfig}
\usepackage{url}

\usepackage[margin=1in]{geometry}
\usepackage{setspace}

\begin{document}
\title{Exploiting TIMSS and PIRLS combined data: multivariate multilevel modelling of student achievement}
\author{Leonardo Grilli\thanks{Department  of Statistics, Informatics, Applications `G. Parenti', University of Florence, Viale Morgagni 59, 50134 Firenze.}\hspace*{1mm},
Fulvia Pennoni\thanks{Department  of Statistics and Quantitative Methods, University of Milano-Bicocca, Via Bicocca degli Arcimboldi 8, 20126 Milano.}\hspace*{1mm}, 
Carla Rampichini\footnotemark[1]\hspace*{1mm} 
and Isabella Romeo\footnotemark[2]} 

\maketitle

\begin{abstract}
We exploit a multivariate multilevel model for the analysis of the Italian sample of the TIMSS\&PIRLS 2011 Combined International Database on fourth grade students. The multivariate approach jointly considers educational achievement on Reading, Mathematics and Science, thus allowing us to test for differential associations of the covariates with the three outcomes, and to estimate the residual correlations between pairs of outcomes at student and class levels. Multilevel modelling allows us to disentangle student and contextual factors affecting achievement. We also account for territorial differences in wealth by means of an index from an external source. The model residuals point out classes with high or low performance. As educational achievement is measured by plausible values, the estimates are obtained through multiple imputation formulas. The results, while confirming the role of traditional student and contextual factors, reveal interesting patterns of achievement in Italian primary schools.

\vspace*{0.5cm} 
\end{abstract}

\section{Introduction}
\label{sec:intro}
The role of large-scale assessment surveys in the public debate about education has dramatically grown since the mid-1980s. Despite the inevitable criticism, international achievement testing has the merit to display the great variability of the educational systems across the world and to shed light on the process underlying the accumulation of the human capital.
As discussed in \cite{Rutkowski:14}, international achievement testing in education has many and ambitious purposes, including the assessment of policies and practices. Indeed, understanding the determinants of achievement in compulsory school is extremely important to design interventions at any level, see among others \cite{Reeve:06} and the references therein.

In this paper we consider the large-scale assessment surveys TIMSS (Trends in International Mathematics and Science Study) and PIRLS (Progress in International Reading Literacy Study) by focusing on the Italian data. These surveys are generally carried out at different years, but in 2011 for the first time the two cycles coincided, thus providing a sample of students with assessments in Reading, Math and Science.
Italy represents an interesting case characterized by a central educational system in a country with marked territorial differences in wealth.

In official reports, for any country the outcomes in Reading, Math and Science are analyzed separately by means of multilevel models \citep{foytechnical:13,Martin:13}.
We propose a  multivariate multilevel approach, where the three scores are treated as a multivariate outcome measured at student level, with students nested within classes.
We demonstrate this approach allows us to gain further insights as we can estimate the residual correlations between pairs of outcomes at both hierarchical levels, which is important to make a comprehensive picture of student achievement and educational effectiveness. Moreover, a multivariate model enables to test whether the relationship of an explanatory variable is identical across outcomes, for example whether gender differences in achievement are the same for Reading and Math.
From a methodological point of view, the multivariate multilevel model is a well established tool \citep{Yang:02}, but its application to the analysis of large-scale assessment data is novel: therefore, we will emphasize the additional insights given by a multivariate approach and we will discuss several issues arising in the implementation, such as the way of handling the imputed scores (known as \emph{plausible values}).

Concerning the Italian data, exploratory analyses show that the three outcomes are highly correlated, in particular at class level. The proportion of variability of the outcomes at class level is relevant, thus calling for an analysis of student characteristics, such as gender, family background, as well as contextual factors, such as school resources and wealth of the surrounding area. To this end we exploit the variables included in the TIMSS\&PIRLS combined dataset, except for wealth which is measured by the gross value-added at the province level, gathered from an external source. 
The considered variables allow us to adjust for prior differences among students and contexts, so to interpret the class level residuals as measures of effectiveness. The term effectiveness is used to denote the unobserved factors at class level affecting student achievement and it should not be interpreted as a causal effect, since students are not randomly allocated to classes and the adjustment for prior differences is necessarily imperfect. With this caveat, the class level residuals can be exploited to point out anomalous situations and further territorial patterns. These features are mainly relevant in those educational systems, like the Italian one, which aim to be egalitarian.

The paper is organized as follows. Section \ref{sec:survey} describes the TIMSS\&PIRLS 2011 survey, whereas Section \ref{sec:italian} focuses on the Italian sample and reports a preliminary analysis. Section \ref{sec:model} outlines the multivariate multilevel model. Section \ref{sec:findings} shows the model selection process and reports the main findings. Section \ref{sec:final} provides concluding remarks and directions for future work. The appendix reports some details on the sampling design and some remarks on the use of the sample weights.

\section{The TIMSS\&PIRLS 2011 survey}
\label{sec:survey}
The large-scale assessment surveys TIMSS and PIRLS are organized by the International Association for the Evaluation of Educational Achievement (IEA). Specifically, TIMSS is an international assessment of mathematics and science achievements at fourth and eighth grades conducted every four years since 1995, whereas PIRLS provides information on trends in reading literacy achievement of fourth grade students every five years since 2001. In 2011 for the first time the TIMSS and PIRLS cycles coincided, enabling the IEA to release the Combined TIMSS\&PIRLS 2011 International Database including fourth grade students responding to both surveys.

In the combined database the two surveys are perfectly comparable since they share the methodological framework and they are administered to the same sample of students. Indeed, IEA released the data as if they were collected by a single survey. In particular, the achievement scores on the three subjects are jointly yielded by a multi-dimensional IRT model preserving the correlation structure.
Furthermore, IEA created additional contextual scales by combining information from the two surveys. For example, the PIRLS scale ``instruction affected by reading resource shortages'' and the TIMSS scales ``instruction affected by mathematics resource shortages'' and ``instruction affected by science resource shortages'' were combined into a new contextual scale labelled ``instruction affected by any resource shortages''.

In TIMSS and PIRLS the students are selected by a complex multi-stage sampling design which is outlined in the Appendix.
The variables are obtained through questionnaires administered to students, their parents, their teachers, and their school principals.
The questionnaires of the two surveys are identical, except for subject-specific issues. For example, questions about teaching math are specific to the TIMSS teacher questionnaire.

Parents completed the home questionnaire with questions about the child (e.g. literacy- and numeracy-centered activities at an early age), the family (e.g. home resources), and the parents themselves (e.g. level of education and employment status). The choice to collect the above information from the parents increases the quality of the derived variables as compared to surveys collecting the information from the student, where the responses may be seriously affected by the socio-economic status of the student \citep{Kreuter:10,Jerrim:14}.

The TIMSS\&PIRLS 2011 database provides achievement results scaled together in a multi-dimensional IRT model in order to preserve the correlation structure across the three achievement scales. Separate achievement scales are produced for Reading, Math, and Science with an international mean of $500$ points and a standard deviation of $100$ points, considering the 32 countries that administered the TIMSS and PIRLS 2011 assessments at the fourth grade \citep{foy:13}.
For each achievement scale, five estimates of the student score, known as \emph{plausible values} are provided. The variability among plausible values 
accounts for the uncertainty inherent in the scale estimation process \citep{Mis:91, martin:12}.
In Section \ref{sec:model} we will exploit the five plausible values for model fitting; however, for simplicity, in the next section we summarizes the data using a single plausible value (the first one) since this yields accurate enough estimates of aggregate quantities, such as class means \citep{Wu:05, Rut:10}.

\section{Preliminary analysis of the Italian sample}
\label{sec:italian}
Our analysis concerns Italy, where $4,200$ students participated to TIMSS and $4,189$ participated to PIRLS. The corresponding TIMSS\&PIRLS 2011 combined dataset includes $4,125$ students who responded to both surveys, thus there are no patterns of missingness in the responses. Note that combining the surveys caused a negligible reduction of the sample size. The students are nested in $239$ classes, which are nested in $202$ schools.  Considering the first plausible values, the average scores for Italy  are (standard deviations in parenthesis):  Reading 525.4 (75.4), Math 502.2 (76.7), and Science 519.0 (78.5).

Italian primary schools belong to a public system: the majority of schools are operated by the state and the other ones must still adhere to common guidelines about study programs. By default a pupil attends the nearest school from home, even if the parents can choose a different school subject to the availability of places. Except in large cities, choosing a different school is uncommon.
Despite the public system, student achievement is markedly different across the nation.
Table \ref{tab:classes} reports sample sizes and summary statistics on achievement for Italy by geographical area, showing a decrease in average scores moving from North to South, whereas the standard deviations among classes have an opposite tendency.
The differences across areas are large, indeed the range of the average score is close to the standard deviation among classes.
The geographical pattern of achievement reflects some well known differences in wealth in Italy. Student achievement is associated with wealth mainly because of the relationship with the socio-economic condition of the area, which is a key factor in the literature on school effectiveness \citep{Hanushek:11}, 
see also Chapter 3 of  \cite{Schuller:07} for an analysis of the multiple role of the formal education on the socio-economic context. Moreover, some services are sustained by the municipalities, thus schools in wealthier areas usually benefit from richer extra-curriculum activities, in addition to better auxiliary services such as canteen and transportation.
Therefore, a fair evaluation of school effectiveness requires to adjust for wealth.

\begin{table}
    \hspace{-2cm}
\caption{\label{tab:classes}    TIMSS\&PIRLS 2011 sample sizes for Italy by geographical area, along with means of the first plausible value (standard deviations among classes in parenthesis) and average Gross Value Added.}
   \vspace{0.5cm} %FP
    \centering
    \fbox{
\scriptsize
\begin{tabular}{l|rrr|rrr|r}
  \emph{Area}     &\multicolumn{3}{c|}{\emph{Sample sizes}}&\multicolumn{3}{c|}{\emph{Average score (sd)}}&GVA\\
                           & Classes &  Teachers   & Students & Read & Math & Scie &    \tabularnewline   \hline
North-West                 & 48  &103&     849&541 (22) & 519 (27)  &539 (28) & 122     \tabularnewline
North-East                 & 49  &103&     920&533 (37) & 509 (40)  &528 (44) & 120     \tabularnewline
Centre                     & 48  & 97&     852&527 (31) & 504 (35)  &522 (38) & 113     \tabularnewline
South                      & 49  & 97&     832&514 (40) & 497 (54)  &508 (53) & 66      \tabularnewline
South-Islands              & 45  & 83&     672&506 (50) & 479 (55)  &494 (62) & 69      \tabularnewline   \hline
Total                      &239  &483&   4,125&524 (39) & 502 (45)  &518 (48) & 100     \tabularnewline
\end{tabular}}
\end{table}

In TIMSS and PIRLS, wealth of the surrounding area is measured through some questions posed to the school principal, but they are of little value due to the subjective nature of the judgement. We thus prefer to rely on an external measure of wealth, namely the \emph{per-capita} Gross Value Added (GVA) at market prices in 2010 \citep{taglia:11}. This index is defined at the province level, which is the finest geographical level available for a nation-wide wealth index. We recognize that GVA neglects the within-province heterogeneity, which can be substantial. Notwithstanding, we argue that GVA is the best available measure for the purpose of adjusting school effectiveness for wealth.

\begin{figure}
    \begin{center}
   \fbox{\begin{tabular}{c|c}
      \includegraphics[width=6cm]{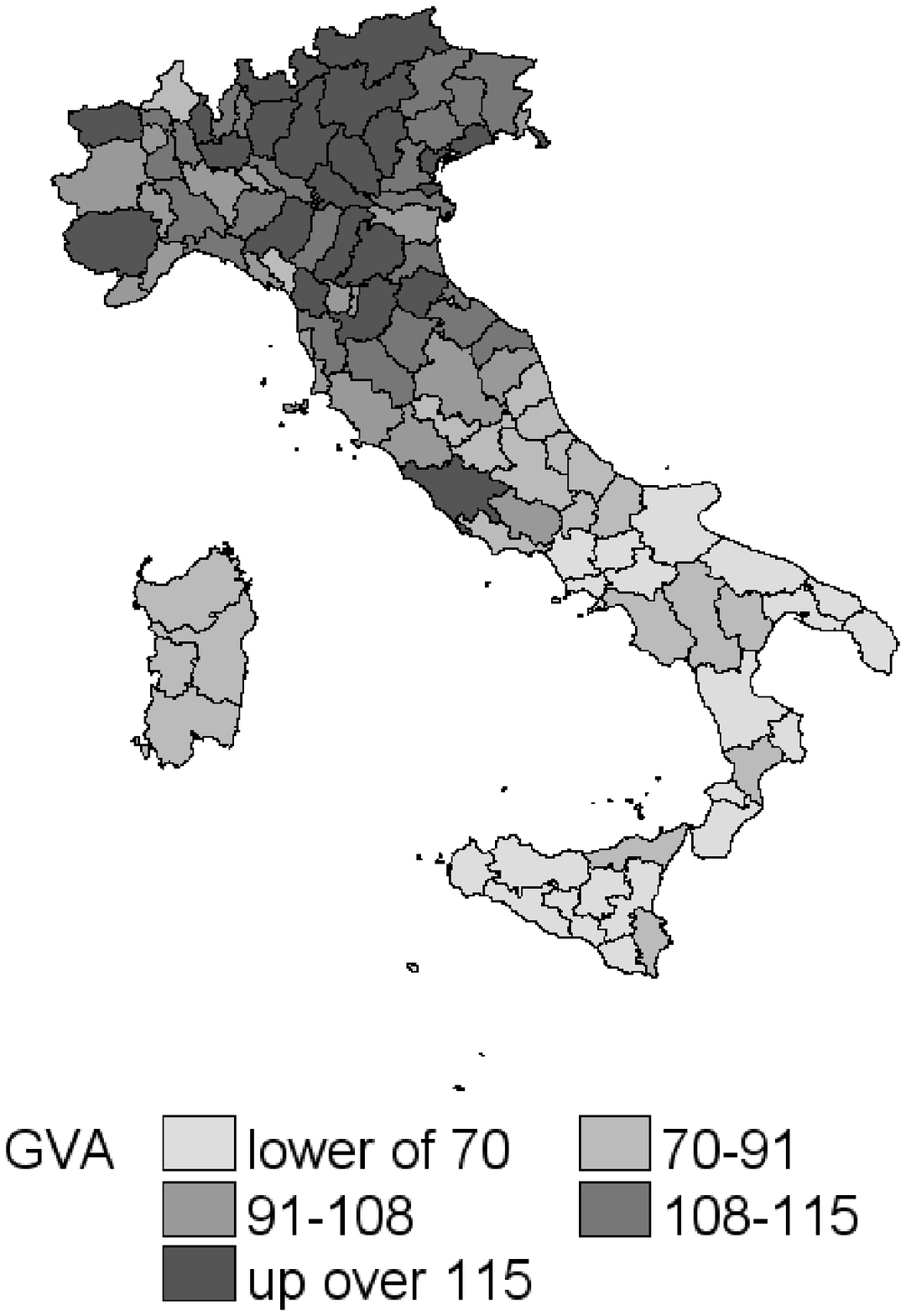} &
     \hspace{0ex} \includegraphics[width=6cm]{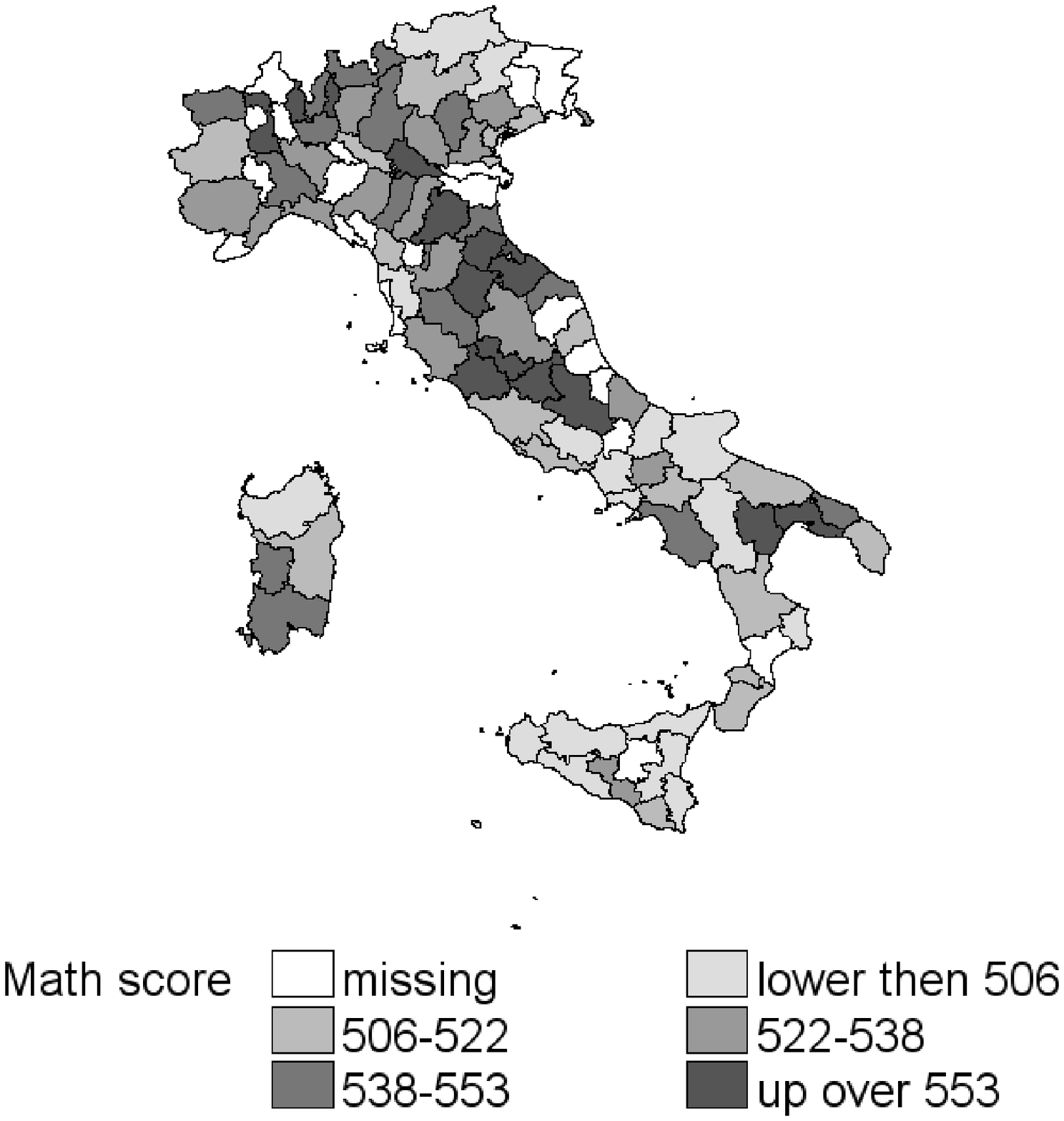}\\
    \end{tabular}}
    %\vspace{-1em}
        \caption{Cartograms by province of the Gross Value Added 2010 (left panel) and the Math average score from TIMSS\&PIRLS 2011 (right panel).}
    \label{fig:cartogramma}
    \end{center}
\end{figure}

\begin{figure}
    \begin{center}
    \makebox{\includegraphics[width=9cm]{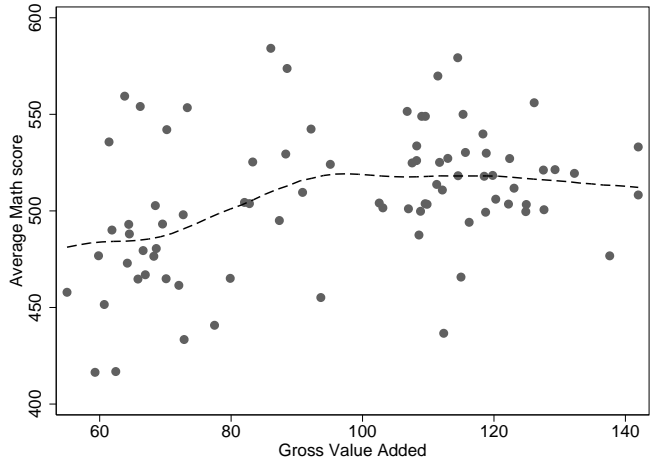}}
     \vspace{-1em}
    \caption{\label{fig:VA_pv} Local polynomial smoothing of the province average of Math score (TIMSS\&PIRLS 2011) as a function of province Gross Value Added (2010).}
        \end{center}
\end{figure}

The index GVA is measured for each of the $110$ Italian provinces, ranging from $55$ to $142$, with $100$ representing the national average.
Figure \ref{fig:cartogramma} shows the patterns at province level of GVA (left panel) and the average score on Math (right panel, where white areas correspond to provinces without sampled schools). Both quantities tend to decrease from North to South, even if the average score on Math is more irregular.
The relationship between Math score and GVA is also represented in Figure \ref{fig:VA_pv} through a local polynomial smoothing: achievement is positively related to wealth, even if the relationship is weak and 
it holds  only for provinces below the Italian average of 100. Reading and Science have similar relationships, therefore their figures are not reported.

In order to adjust the achievement scores for student and contextual factors, we consider a subset of the variables available in the combined dataset. The choice of this subset is driven by theoretical arguments,
as suggested by \cite{Hanushek:11}, as well as exploratory data analysis and previous studies, (see in particular the technical Appendix B of the TIMSS\&PIRLS 2011 Report in \cite{foytechnical:13}).
Descriptive statistics of the selected covariates are shown in Table \ref{tab:summary}.
The first column reports the sample sizes at the relevant hierarchical levels (student, teacher, class, school), whereas the last column reports the number of observations for each variable: in most cases there are missing values, though their percentage is small (at most 8.6\%).
\begin{table}
\caption{\label{tab:summary} Summary statistics of selected variables by hierarchical level (sample size in parenthesis, TIMSS\&PIRLS 2011, Italy).}
\vspace{0.5cm}
\fbox{%
\scriptsize
\begin{tabular}{llrrrrr}
\multicolumn{1}{l}{{Level}} & \multicolumn{1}{l}{{Variable}} & {Mean}
 & {Std. Dev.}& {Min.} &  {Max.} & {N} \tabularnewline
\hline
\multicolumn{1}{c}{\multirow{5}[2]{*}{Student (4,125)}} & Female & 0.51 & - & 0 & 1 & 4,125 \tabularnewline
& Pre-school                        & 0.75 & - & 0 & 1 & 3,826 \tabularnewline
& Language at home not Italian      & 0.21 & - & 0 & 1 & 4,083 \tabularnewline
& Home resources for learning       & 9.72 & 1.55 & 3.41 & 15.29 & 3,770 \tabularnewline
& Early literacy/numeracy tasks     & 9.24 & 1.60 & 3.65 & 12.90 & 3,846 \tabularnewline
\hline
& Woman                             & 0.96 & - & 0 & 1 & 458 \tabularnewline
& Age                               &  &  &  &  &  458\tabularnewline
\multicolumn{1}{c}{\multirow{12}[2]{*}{Teacher (483)}} & \hspace{2ex} $<40$ &0.15&&&&    \tabularnewline
& \hspace{2ex} $40-49$                          &0.38   &&&&    \tabularnewline
& \hspace{2ex} $\ge 50$                         &0.47   &&&&    \tabularnewline
& Degree & 0.35 & - & 0 & 1 & 455   \tabularnewline
& Years of teaching & 23.34 & 10.29 & 1 & 42 & 446 \tabularnewline
& Subject(s) taught             &  &  &  &  &483  \tabularnewline
& \hspace{2ex} Reading only                      &0.35   &&&&    \tabularnewline
& \hspace{2ex} Mathematics and Science           &0.28   &&&&    \tabularnewline
& \hspace{2ex} Mathematics only                  &0.12   &&&&    \tabularnewline
& \hspace{2ex} Science only                      &0.10   &&&&    \tabularnewline
& \hspace{2ex} Mathematics, Science and Reading  &0.06   &&&&    \tabularnewline
& \hspace{2ex} Science and Reading               &0.05   &&&&    \tabularnewline
& \hspace{2ex} Mathematics and Reading           &0.03   &&&&    \tabularnewline
\hline
\multicolumn{1}{c}{\multirow{5}[1]{*}{Class (239)}} &
  Mean of `Female'                  & 0.51 & 0.13 & 0 & 1 & 239 \tabularnewline
& Mean of `Pre-school'              & 0.69 & 0.17 & 0 & 1 & 239 \tabularnewline
& Mean of `Language at home not Italian' & 0.23 & 0.17 & 0 & 1 & 239 \tabularnewline
& Mean of `Home resources for learning' & 9.64 & 0.83 & 7.17 & 11.90 & 238 \tabularnewline
& Mean of `Early literacy/numeracy tasks' & 9.24 & 0.51 & 7.58 & 10.89 & 238 \tabularnewline
\hline
\multicolumn{1}{c}{\multirow{9}[2]{*}{School (202)}}
& Adequate environment and resources & 9.62 & 1.07 & 7.13 & 13.62 & 201 \tabularnewline
& School is safe and orderly & 9.41 & 0.88 & 7.36 & 12.37 & 202 \tabularnewline
& School with Italian students $>$ 90\% & 0.65 & 0.48 & 0 & 1 & 196 \tabularnewline
& Less than 10\% students with low SES & 0.39 & 0.49 & 0 & 1 & 191 \tabularnewline
& School is located in a big area & 0.34 & 0.48 & 0 & 1 & 198 \tabularnewline
& In the area live more then 50.000 people & 0.28 & 0.45 & 0 & 1 & 196 \tabularnewline
& Six days of school per week & 0.47 & 0.50 & 0 & 1 & 198 \tabularnewline
\end{tabular}}
\end{table}
At student level, we include dummy variables for gender (1 if female), pre-school (1 if the student attended at least 3 years) and language spoken at home (1 if not Italian). Furthermore, two home background questionnaire scales from TIMSS\&PIRLS 2011 are used to describe the student home environment: \emph{Home Resources for Learning} and \emph{Early Literacy/Numeracy Tasks}, described in detail by \cite{Martin:13}. In summary, \emph{Home Resources for Learning} is derived from items on the number of books and study supports available at home and parents' levels of education and occupation; on the other hand, \emph{Early Literacy/Numeracy Tasks} is the student average score on two scales derived from the parents' responses on how well their child could do some early literacy and numeracy activities when beginning primary school.

At teacher level we consider gender (1 if woman), age group, education (1 if the teacher has a degree) and years of teaching.
Note that the average number of years of teaching (23.34) is rather high: in fact, in the past it was common to start teaching quite early since the requirement was the diploma from a special purpose high school (the university degree was not necessary, indeed only 35\% have a degree); moreover, in recent years the average age of teachers increased due to a low turnover.
Any teacher is in charge of one or more subjects. Table \ref{tab:summary} shows that Reading usually has a specific teacher, whereas Math and Science are often taught by the same teacher, even if also other combinations of teachers and subjects are possible. 

The class variables in Table \ref{tab:summary} are defined as averages of the corresponding student level covariates. At school level, \emph{Adequate environment and resources} and \emph{School is safe and orderly} are contextual scales \citep{Martin:13}, while the other school variables in Table \ref{tab:summary} are directly based on the answers of school principals.

\section{Model specification}
\label{sec:model}
We outline the notation for the multivariate multilevel model \citep{Yang:02, Goldstein:11, Sni:12}. Let $Y_{mij}$ be the score on the $m$-th outcome for the $i$-th student of the $j$-th class, with $m=1,2,3$ (1: Reading, 2: Math, 3: Science), $i=1,\ldots,n_j$, $j=1,\ldots, J$. The number of students of the $j$-th class is denoted with $n_j$, whereas the total number of students is denoted with $N = \sum _{j=1} ^{J} n_j$. The Italian sample of the TIMSS\&PIRLS 2011 Combined Dataset includes $N=4,125$ students nested into $J=239$ classes.

Official reports \citep{foytechnical:13} consider students as level 1 units and schools as level 2 units. However, in our model the level 2 is represented by classes since several factors act at the class level (e.g. the peer effects), thus merging different classes would obscure some sources of variability.
The schools could be considered as a third hierarchical level with their own random effects. However, in the Italian sample only 37 schools out of 202 have more than one class, thus there is little information to estimate also school level variances.
Nevertheless, the school characteristics are included in the model as covariates and the correlation between classes of the same school is accounted by robust standard errors for clustered observations \citep{Rabe:06}.

We specify the following multivariate two-level model for outcome $m$ of student $i$ in class $j$:
\begin{equation}\label{eq:model}
    Y_{mij}=\alpha_{m}+\bm\beta_{m}' \mathbf{x}_{mij}+\bm\gamma_{m}' \mathbf{w}_{mj}+u_{mj}+e_{mij}
\end{equation}
where $\mathbf{x}_{mij}$ is the vector of student level covariates, $\mathbf{w}_{mj}$ is the vector of class level covariates, also including covariates at higher level, e.g. school or province. All the vectors have the outcome index $m$ since they can include outcome-specific covariates, such as the characteristics of the teacher.
Student level errors $e_{mij}$ are assumed independent across students, and class level errors $u_{mj}$ are assumed independent across classes. The errors $e_{mij}$ are independent from the errors $u_{mj}$.
In model (\ref{eq:model}) student level errors $\mathbf{e}_{ij}'=(e_{1ij},e_{2ij},e_{3ij})$ are assumed to be multivariate normal with zero means and covariance matrix
\begin{equation}\label{eq:vare}
Var(\mathbf{e}_{ij})=\bm\Sigma=\left(
  \begin{array}{ccc}
    \sigma_1^2 & \sigma_{12} & \sigma_{13}\\
     & \sigma_2^2 & \sigma_{23}\\
     & & \sigma_3^2 \\
  \end{array}
\right)
\end{equation}
whereas class level errors $\mathbf{u}_{j}'=(u_{1j}, u_{2j},u_{3j})$ are assumed to be multivariate normal with zero means and covariance matrix
\begin{equation}\label{eq:varu}
Var(\mathbf{u}_{j})=\bm T=\left(
  \begin{array}{ccc}
    \tau_1^2 & \tau_{12} & \tau_{13}\\
     & \tau_2^2 & \tau_{23}\\
     & & \tau_3^2\\
  \end{array}
\right).
\end{equation}
Therefore, the response vector $\mathbf{Y}_{ij}=(\mathit{Y}_{1ij},\mathit{Y}_{2ij},\mathit{Y}_{3ij})'$ has residual covariance matrix given by the sum of  $\bm\Sigma$ and $\bm T$.
The class level error $u_{mj}$ represents unobserved factors at class level for outcome $m$, including the teacher effect. Note that the school level is omitted, thus $u_{mj}$  accounts also for school unobserved factors.

The assumptions on the model errors listed above are the standard ones \citep{Grilli:15}. Alternatively, we considered several specifications with heteroscedastic random effects, for example to account for differential variability across geographical areas \citep{Sani:11}. However, none of the heteroscedastic specifications significantly improved the model fit, which is not surprising in the light of the complexity of the considered model, where the random effects at the class level are characterized by a $3 \times 3$ covariance matrix.
A peculiar heteroscedastic specification that is worth to be mentioned is related to the fact that in some classes one teacher teaches all the three subjects under consideration, while in other classes there are different teachers depending on the subject (see Table \ref{tab:summary}).

The sample correlations between outcomes are slightly higher in classes with a single teacher. This pattern can be accounted by a model with a covariance matrix at the class level depending on the kind of teacher allocation. We tried to estimate separate covariance matrices for classes with a single teacher and classes with multiple teachers, but
we could not gain  significant improvements in the model fit. Therefore, we proceed with the multivariate model with a unique class level covariance matrix.

\section{Model selection and results}
\label{sec:findings}
The analysis is based on the multivariate two-level model of equations (\ref{eq:model})-(\ref{eq:varu}) fitted by maximum likelihood.
In order to account for the variability induced by plausible values,
estimation is performed separately for each of the five plausible values and then the results are combined by using Multiple Imputation (MI) formulas \citep{Rubin:02, Schafer:03}.
These formulas yield correct standard errors accounting for both the variability in imputing the scores and the variability in estimating the model parameters.
The analysis is carried out by using the \texttt{mixed} and \texttt{mi} commands of Stata \citep{Stata:13}.

The estimation sample consists of 3,741 students and 237 classes.
Indeed, 384 students (9.3\%) and 2 classes (0.8\%) have been excluded due to missing values in the covariates (see Table \ref{tab:summary}).
In our application, missing values are rare at class and school levels.
At the student level, the covariates with the higher percentage of missing values are those derived from the parents questionnaire, namely \emph{Pre-school} ($7.2\%$), \emph{Home resources for learning} ($8.6\%$) and \emph{Early literacy/numeracy tasks} ($6.8\%$).

An alternative to deleting records with missing values is represented by multiple imputation methods, which have recently been extended to complex multilevel settings \citep{Goldstein:14}. In the framework of large-scale assessment data, \cite{Bouhlila:13} applied multiple imputation with chained equations to TIMSS 2007 data, while \cite{foytechnical:13} applied single imputation to TIMSS 2011 data.  \cite{Weirich:14} conducted a simulation study to evaluate the performance of imputation methods to handle missing background variables in the IRT model for generating the plausible values.
In the data we dispose, missing values mostly concern the covariates derived from the parents questionnaire. Unfortunately, those covariates tend to be missing altogether, thus it is difficult to specify an effective imputation model.
The missing mechanism seems to be related to the family background, so deleting students with missing values may lead to some bias in the estimates. This kind of bias is likely to remain still after imputing missing values \citep{Rubin:02}. Weighting the costs and benefits, we decided to proceed without imputation.

In the following we first show the results from the null model, i.e. without covariates, then we outline the model selection strategy and finally we illustrate the results from the final model.

\subsection{Results from the null model}
In order to explore the correlation structure of the three outcomes,
we first fit the null model, which has 3 parameters for the means and 12 parameters for the variances and covariances (6 at student level and 6 at class level).
Table \ref{tab:corr} summarizes the results of the null model in terms of correlation matrices and between class proportions of variances and covariances after the application of MI formulas.
The \emph{within class} and \emph{between class} correlation matrices are derived from the corresponding covariance matrices $\bm \Sigma$ and $\bm T$ of equations (\ref{eq:vare}) and (\ref{eq:varu}), whereas the total correlation matrix is derived from the total covariance matrix ($\bm \Sigma + \bm T$).
Table \ref{tab:corr} shows that the three scores are highly correlated, in particular at class level.

The rightmost matrix in Table \ref{tab:corr} reports the percentage of variances and covariances at the class level, namely each element of $\bm T$ is divided by the corresponding element of ($\bm \Sigma + \bm T$). For example, the percentage of class level variance for Reading is $100 \times \hat{\tau}^2_{1}/(\hat{\sigma}^2_{1}+\hat{\tau}^2_{1}) = 19.8$, which is also known as ICC (Intraclass Correlation Coefficient).
Note that Reading is the subject with the lowest ICC, maybe because it is most influenced by student background  characteristics.

The proportion of variability of the scores at class level is relevant, thus calling for an analysis of contextual factors.
To this end, we select the covariates summarized in Table \ref{tab:summary} as explained in the following subsection.
\begin{table}
\caption{\label{tab:corr} Multivariate multilevel model: correlation matrix decomposition from the null model, MI combined estimates (TIMSS\&PIRLS 2011, Italy).}
\vspace{0.5cm}
\centering
\renewcommand{\baselinestretch}{1}
\fbox{
\scriptsize
\begin{tabular}
{>{\raggedright}p{1cm}|>{\raggedright}p{0.6cm}>{\raggedleft}p{0.6cm}>{\raggedleft}p{0.6cm}|>{\raggedleft}p{0.6cm}>{\raggedleft}p{0.6cm}>{\raggedleft}p{0.6cm}|>{\raggedleft}p{0.6cm}>{\raggedleft}p{0.6cm}>{\raggedleft}p{0.6cm}|>{\raggedleft}p{0.6cm}>{\raggedleft}p{0.6cm}>{\raggedleft}p{0.6cm}}
&\multicolumn{9}{c|}{\emph{Correlations}}&\multicolumn{3}{c}{\% \emph{Between class}}\\ \cline{2-10}
&\multicolumn{3}{c|}{\emph{Within class}}&\multicolumn{3}{c|}{\emph{Between class}}&\multicolumn{3}{c|}{\emph{Total}}&\multicolumn{3}{c}{\emph{of (co)variances}}\\ \cline{2-13}
\emph{Subject }    & Read & Math & Scie& Read & Math & Scie& Read & Math & Scie& Read & Math & Scie \tabularnewline \hline
Reading               &1.00&&          & 1.00&&                &1.00&&                &19.8&&                \tabularnewline
Math               &0.71&1.00&      & 0.93&1.00&            &0.76&1.00&            &29.5&28.8&            \tabularnewline
Science           &0.81&0.74&1.00  & 0.97&0.98&1.00        &0.85&0.81&1.00        &28.2&35.0&29.4        \tabularnewline
\end{tabular}}
\end{table}
\normalsize

\subsection{Model selection}
The model selection process in principle involves to fit the multivariate multilevel model repeatedly, each time combining the estimates with MI formulas.
In order to speed up the selection process, we adopt two simplifications: (\emph{i}) the outcomes are analyzed separately with univariate multilevel models, retaining covariates being significant in at least one of the univariate models, and (\emph{ii}) estimation is carried out using only the first plausible value.
Using a single plausible value gives underestimated standard errors, implying a conservative selection of the covariates.

In order to enhance the interpretability of the intercept, we centered continuous covariates at their sample grand means, except for GVA which is centered at 100 (Italian average). As shown in Table \ref{tab:summary}, we consider as covariates all the class level means of the student covariates; however, contrary to official reports \citep{foytechnical:13}, we do not center student level covariates at their class level means.
The covariates have been added according to the hierarchy: student, class, school, province.

Table \ref{tab:models} reports the models selected at the end of each hierarchical step.
All the considered models have 12 variance-covariance parameters: 6 for the within class covariance matrix in equation (\ref{eq:vare}) and 6 for the between class covariance matrix in equation (\ref{eq:varu}).
Teacher variables have been added to model \emph{M1}, but none are significant.
Note that models \emph{M2} and \emph{M3} have the same number of parameters since, after the inclusion of GVA, the class mean of \emph{Early literacy/numeracy tasks} is no more significant.
\begin{table}
\caption{\label{tab:models} Multivariate multilevel model: main steps of the model selection process, first plausible value  (TIMSS\&PIRLS 2011, Italy).}
\vspace{0.5cm}
\centering
\fbox{%
\scriptsize
\begin{tabular}{lccl}
  Model &     n.par. &   $logL$      & Significant covariates (on at least one subject)\\    \hline
    \emph{M0}: null                &   15 &  -59,625.44 &            \\
    \emph{M1}: student  covariates &   30 &  -59,119.09 & {\multirow{3}{*}{\parbox{6cm}{Female, Pre-school, Language spoken at home is not Italian, Home resources for learning, Early literacy/numeracy tasks }}}\\
    &&&\\
    &&&\\
{\multirow{2}{*}{\parbox{4cm}{\emph{M2}: student and class/school covariates}}} &     36 & -59,109.75 &                 {\multirow{3}{*}{\parbox{6cm}{M1 covariates +  Class average Early literacy/numeracy tasks,  School adequate environment and resources}}}\\
    &&&\\
    &&&\\
{\multirow{2}{*}{\parbox{4cm}{\emph{M3}: student, class/school and province  covariates}}}  & 36 & -59,106.25 & {\multirow{2}{*}{\parbox{6cm}{M1 covariates + School adequate environment and resources,  Gross Value Added by province}}}\\
    &&&\\
\end{tabular}}
\end{table}

\subsection{Results from the final model}
The results from the final model are obtained by fitting model \emph{M3} of Table \ref{tab:models} separately for each plausible value and then combining the estimates through MI formulas.
Table \ref{tab:finalmodel} reports the estimates of regression coefficients and variance-covariance parameters alongside with robust standard errors. With reference to the considered covariates summarized in Table \ref{tab:summary}, all the student level covariates are significant, but the corresponding class means are not. None of the teacher covariates are significant.
At school level, the only significant variable is \emph{Adequate environment and resources}. At province level, GVA is significant.

The last column of Table \ref{tab:finalmodel} reports the $p$-value of the  test \emph{F} for the equality of the regression coefficients across the three outcomes. For example, for the $s$-th student level covariate the null hypothesis is $H_0: \beta_{1s}=\beta_{2s}=\beta_{3s}$. The test, which is feasible only in a multivariate model, is performed with the command \texttt{mi testtr} of Stata, implementing formula (1.17) of \cite{Li:91}.
Interestingly, for all the contextual covariates the magnitude of the association with the three outcomes is similar, while the student level covariates show different relationships with the outcomes, except for pre-school.
Also note that family background covariates have a similar association with Reading and Science, as opposed to Math, therefore the abilities required for Science seem to be closer to those for Reading than to those for Math.
Likely, this is a consequence of the fact that, in Italian primary schools, the approach to Science is mainly qualitative, thus reading ability is more important than math ability.
\begin{table}
\caption{\label{tab:finalmodel} Multivariate multilevel model: parameter estimates  and robust standard errors from the final model, MI combined results (TIMSS\&PIRLS 2011, Italy).}
\vspace{0.5cm}
\centering
\fbox{%
\scriptsize
\begin{tabular}{lrrrrrrr}
                   & \multicolumn{2}{c}{Read} & \multicolumn{2}{c}{Math} & \multicolumn{2}{c}{Science} & Test \emph{F}$\dag$\tabularnewline
                   &      Coef. & s.e. &      Coef. &  s.e. &      Coef. & s.e.&\emph{p}-value\tabularnewline
\hline
\emph{Regression coefficients}&&&&&&\tabularnewline
\hspace{2ex}Intercept  &    531.73  &       3.57 &514.99 &       4.25 &      531.47 &       3.92 & 0.0006 \tabularnewline
\emph{Student covariates}&&&&&&&\tabularnewline
\hspace{2ex} Female   &       2.92 &       2.41 &-11.96 &       3.05 &      -10.64 &       2.28&0.0000 \tabularnewline
\hspace{2ex} Language at home is not Italian &-22.57 & 3.12 &-14.94 &   3.27 & -23.74 & 3.53 & 0.0161 \tabularnewline
\hspace{2ex} Pre-school     &      8.85 &       3.01 &  8.46 &    2.51 &       10.91 &  3.15 & 0.6386 \tabularnewline
\hspace{2ex} Home resources for learning &   14.04 &   0.84 & 10.64 &  0.84 &  13.23 & 0.93 &0.0009\tabularnewline
\hspace{2ex} Early literacy/numeracy tasks & 7.24 &    0.77 & 10.07 &  0.76 &    6.53 &   0.83 &0.0051\tabularnewline
\emph{School covariates}&&&&&&&\tabularnewline
\hspace{2ex} Adequate environment \& resources &5.28 &1.92 &  8.61 &  3.19 &  7.00 &  2.96 & 0.1950 \tabularnewline
\emph{Province covariates}&&&&&&&\tabularnewline
\hspace{2ex} GVA (below 100)  &  0.45 &  0.15 &  0.48 & 0.21 &  0.55 &  0.20 &0.3983\tabularnewline  \hline
\multicolumn{1}{l}{\emph{Between-classes covariance matrix}}&&&&&&&\tabularnewline
\hspace{2ex} Variances: \hspace{9ex} $\tau_m^2$       &725.7 &192.0&1332.3 & 225.1&1274.1 &  262.1&\tabularnewline
\hspace{2ex} Cov(Read,Math): \hspace{0.7ex} $\tau_{12}$     &915.2 &195.9&       &      &    &    &\tabularnewline
\hspace{2ex} Cov(Math,Scie): \hspace{1.8ex} $\tau_{23}$ &   &   &  1266.1 & 234.7&      &         &\tabularnewline
\hspace{2ex} Cov(Read,Scie): \hspace{2.9ex}$\tau_{13}$  &   &   &         &      &931.6 &   221.1 &\tabularnewline
\multicolumn{1}{l}{\emph{Within-classes covariance matrix}}&&&&&&&\tabularnewline
\hspace{2ex} Variances: \hspace{9ex} $\sigma_m^2$    &3716.1&   101.9&3500.1 & 120.8 &3471.7&  132.6  &\tabularnewline
\hspace{2ex} Cov(Read,Math): \hspace{0.7ex} $\sigma_{12}$   &2400.9 & 86.0  &       &      &       &  &\tabularnewline
\hspace{2ex} Cov(Math,Scie): \hspace{1.8ex} $\sigma_{23}$  &      &     &2452.3& 91.6 &       &       &\tabularnewline
\hspace{2ex} Cov(Read,Scie): \hspace{2.4ex} $\sigma_{13}$  &      &     &       &     &2757.4 & 105.9 &\tabularnewline
\multicolumn{8}{l}{$\dag$ for the equality of regression coefficients among the three outcomes.}
\end{tabular}}
\end{table}
\normalsize

The intercepts in Table \ref{tab:finalmodel} represent the average scores for the baseline student: male, language spoken at home is Italian, not attended pre-school, and all the other covariates set at mean values. The performance of the baseline student is beyond the international mean of $500$ in all the considered outcomes, though the average score in Math is substantially lower than the average scores in Reading and Science. According to the test \emph{F}, this difference is significant.

All the regression coefficients have the expected signs and are significant for all the considered outcomes, except for being female, which has a negative association for Math and Science, but no significant effect for Reading.
Students  from  families not speaking Italian at home have a lower performance, especially in Reading and Science. Students who attended pre-school for at least three years have a better performance, with no significant difference among the three outcomes. The two home background questionnaire scales have a positive association on student achievement. However, \emph{Home Resources for Learning} (including number of books at home and education level and employment status of parents) has a greater association on Reading and Science, while \emph{Early Literacy/Numeracy Tasks} (measuring how well the child could do several early literacy and numeracy activities when beginning primary school) has a stronger effect on Math. Thus, achievements in Reading and Science are more related to cultural and socio-economic factors of the family, while achievement in Math is more related to specific activities in early childhood.
At school level, disposing of an \emph{Adequate Environment and Resources} helps to reach a higher score, with no significant difference across outcomes.

The socio-economic context of the province where the school is located is measured by the GVA index. On the basis of the local polynomial smooth of Figure \ref{fig:VA_pv}, the influence of GVA is modelled by a linear spline with a single knot in $100$ (the national average). Consistently with the relationship highlighted in Figure \ref{fig:VA_pv}, the line for GVA$<100$ has a significant positive slope, whereas the line for GVA$>100$ is nearly flat and the slope is not significantly different from zero. Therefore, we constrain to zero the slope of the second line of the spline (i.e. GVA$>100$), so that wealth affects student achievement only in provinces with GVA below the national average. The influence of GVA is similar across outcomes and it amounts to about half point in the score for each point in the index. For example, the influence of GVA on the achievement scores is about $-22.5$ points for the province with the lowest value of GVA ($55$).

The proportions of variance explained by the final model with respect to the null model are higher at class level. Indeed, the within-class variances reduce by about $15$\% for the three outcomes, whereas the between-class variances reduce by $33$\% for Reading, $20$\% for Math and $26$\% for Science. The reduction of the between-class variances is due to the contextual variables at school and province levels and to the compositional effects of student background covariates. Such compositional effects capture cultural and socio-economic factors that are more related to the achievement in Reading, whose class level residual variance shows the greater reduction.
Even if the reduction of variances is stronger at the class level, the residual ICC's derived from the estimated variances of Table \ref{tab:finalmodel} are quite high, specifically $16.3$\% for Reading, $27.6$\% for Math and $26.9$\% for Science. These values point out the existence of unobserved relevant class level factors. The correlations among outcomes are similar to those observed in the null model, reported in Table \ref{tab:corr}. In particular, the estimated correlations among the class level errors of the three outcomes are very high, namely at least $0.93$.

\subsection{Residual analysis}
In the multivariate multilevel model in (\ref{eq:model}) the level 2 error $u_{mj}$ is a class level random effect representing the contribution of class $j$ to the achievement of students in subject $m$. This contribution can be interpreted in terms of effectiveness since the model adjusts for differences in student characteristics and contextual factors. Given the high residual correlations among the random effects of the three outcomes (Reading, Math and Science), the residual analysis can focus on a single outcome without any relevant loss of information. In the following, we illustrate the analysis with reference to Math.

In principle, the residual analysis could be carried out by combining the results derived from the models fitted on the five plausible values. However, this is not essential since quantities at class level are little affected by the variability induced by the plausible values. Therefore, to simplify the analysis, we consider the level 2 residuals of the model based on the first plausible value. Once the model has been fitted via maximum likelihood, the level 2 errors $u_{mj}$ are predicted by the level 2 residuals $\hat{u}_{mj}$, obtained as usual by the Empirical Bayes (EB) method
 \citep{Goldstein:11, Sni:12}.
\begin{figure}
    \begin{center}
            \makebox{\includegraphics[height=7cm]{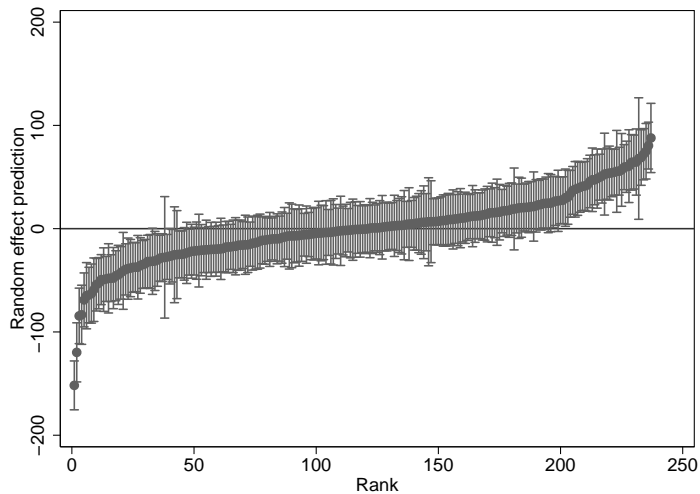}}
            \vspace{-1em}
    \caption{\label{fig:caterpillar} Empirical Bayes predictions of the random effects for Math with 95\% confidence intervals, first plausible value. TIMSS\&PIRLS 2011, Italy.}
    \end{center}
\end{figure}
In order to compare classes in terms of effectiveness, Figure \ref{fig:caterpillar} shows a caterpillar plot where EB residuals are reported in increasing order and endowed with 95\% confidence intervals defined as $\pm 1.96$ times the comparative standard error.
Classes whose confidence interval do not intersect zero have a degree of effectiveness significantly different from the population mean. Specifically, classes with an interval above zero are \emph{good} since the student average achievement is significantly higher than the level expected on the basis of their covariates. Analogously, classes with an interval below zero are \emph{poor}. Figure \ref{fig:caterpillar} shows that, out of $237$ classes, $41$ are \emph{good} and $41$ are \emph{poor}.

\begin{figure}
    \begin{center}
    \makebox{\includegraphics[height=7cm]{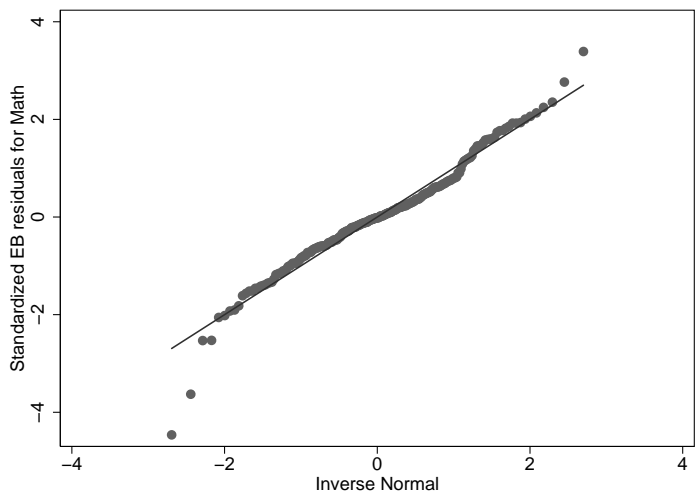}}
    \vspace{-1em}
    \caption{\label{fig:qqplot} Normal probability plot of standardized Empirical Bayes residuals for Math, first plausible value. TIMSS\&PIRLS 2011, Italy.}
        \end{center}
\end{figure}

The model accounts for differences in wealth across provinces by means of the GVA index, thus the residuals may reveal further territorial differences not captured by GVA.
Table \ref{tab:value_added} reports the proportions of \emph{good} and \emph{poor} classes by geographical area based on level 2 residuals: in North-West \emph{good} classes prevail on \emph{poor} classes, while in the Centre the pattern is reversed. This points out a residual territorial influence on mean achievement beyond GVA. This influence could be accounted by  geographical dummies in the fixed part of the model, but their coefficients turn out to be not significant.
In the two Southern areas, the proportions of \emph{good} and \emph{poor} schools are higher than in the rest of Italy. This confirms that schools in Southern regions have a higher variability in effectiveness, as found by \cite{Sani:11} using national standardized test data collected by the Italian Institute for the Evaluation of the Educational System (INVALSI). As mentioned in Section \ref{sec:model}, such differential variability could be modelled through heteroskedastic random effects, but in the present application there is no significant improvement in model fit.
\begin{table}
\caption{Proportions of \emph{good} and \emph{poor} classes based on EB residuals from the final model $M3$. TIMSS\&PIRLS 2011, Italy.
\label{tab:value_added}}
\vspace{0.5cm}
\centering
\fbox{
\scriptsize
\begin{tabular}{lcrr}
Area              & Classes& \multicolumn{2}{c}{Proportion of classes}  \\
                  &        &  \emph{good} &  \emph{poor}            \\ \hline
North-West        &  48    &   0.167&        0.063 \\
 North-East       &  49    &   0.102&        0.082 \\
    Centre        &  47    &   0.085&        0.234 \\
     South        &  49    &   0.286&        0.245 \\
South and Islands &  44    &   0.227&        0.250 \\ \hline
Italy             & 237    &  0.173 &        0.173 \\
\end{tabular}}
\end{table}
\normalsize
In order to inspect the distribution of the random effects, EB residuals are standardized with diagnostic standard errors and depicted in the normal probability plot of Figure \ref{fig:qqplot}. Caution is needed since misspecification may be difficult to detect from the residuals \citep{McCulloch:11}; anyway, the plot in Figure \ref{fig:qqplot} is reassuring as it does not show serious deviations from normality. In addition, there are only three outlying classes (standardized residual outside the $\pm 3$ interval).

\vspace{1cm}

\section{Final remarks}
\label{sec:final}
We have carried out a secondary data analysis of the Italian sample of the TIMSS\&PIRLS 2011 Combined International Database. This database provides an opportunity to perform, for the first time with TIMSS and PIRLS surveys, a joint analysis of achievement in Reading, Math and Science for fourth grade students. The analysis relies on a multivariate multilevel model, thus accounting for both the multivariate nature of the outcome and the hierarchical structure of the data.

The additional findings allowed by the multivariate approach are twofold.
First, by estimating the correlations among the outcomes, we found that achievements in Reading, Math and Science are highly correlated at both student level and class level, even after adjusting for individual and contextual factors. At class level, correlations are so high that in terms of effectiveness the three outcomes are essentially indistinguishable.
Second, by testing for differences in the regression coefficients, we found that females have a lower performance in Math and Science, but not in Reading, and student background covariates have similar influence on Reading and Science, as opposed to Math; on the other hand, contextual covariates have similar influence on the three outcomes.

A further peculiarity of our analysis lies in the use of the per-capita Gross Value Added at province level (GVA) as an external indicator accounting for territorial differences in wealth. The relationship between student achievement and GVA is well represented through a spline: it is found that student achievement is positively related to wealth for provinces below the national average, with no significant relationship for provinces above the national average. It is worth to note that GVA effectively replaced the dummy variables for the geographical areas, yielding a parsimonious and more interpretable model.
From the latter we carried out the analysis of class level residuals which allowed us to identify  few classes with extremely high or low effectiveness and to investigate further patterns not described by the model, such as the higher variability in effectiveness among classes in the Southern regions.

In general, the assessment of school effectiveness should adjust for prior achievement, for example the level of achievement at the beginning of the school cycle. TIMSS\&PIRLS data do not contain any direct measure of prior achievement, but this is a minor limitation since (\textit{i}) the primary school is the first compulsory cycle, thus the programs start with the basics of each discipline, and (\textit{ii}) the data contain some variables related to the skills of the pupil acquired before the primary school, in particular we exploited the indicator for having attended at least 3 years of pre-school and the scale \emph{Early Literacy/Numeracy Tasks}. The latter 
this covariate is intended to be a baseline ability control, even if it is also related to the home environment.

The scaling methodology of TIMSS\&PIRLS is similar to that of PISA, which has been criticized by several scholars \citep{Kirsch:02, Goldstein:04, Kreiner:14}. However, we argue that scaling issues have a limited impact on our results since we consider a given grade at a fixed time in a single country. In particular, we do not make comparisons across countries.
Notwithstanding, there remain concerns about the adequacy of the model used to generate the plausible values, namely the imputation model. In fact, the TIMSS\&PIRLS documentation explains that the model is conditioned on a subset of the principal components of the background variables, in addition to a few `primary' variables such as student gender.
The imputation model is not a multilevel one, thus the standard MI formulas do not necessarily produce unbiased estimates of standard errors. 
In general, there are no details on model specification and fit, thus it is
not possible to judge the adequacy of the imputation model.

Despite their richness, TIMSS\&PIRLS data are collected by a cross-section design, thus preventing to study the dynamics the of achievement process. To overcome this limitation, the Italian Institute for the Evaluation of the Educational System (INVALSI) is currently planning longitudinal surveys. The potentialities of longitudinal data are illustrated by \cite{Bartolucci:11}, who exploited a pilot survey for the Lombardy region to fit a multilevel latent Markov model addressing interesting research questions on achievement progress.

\section*{Acknowledgements}
The research has been supported by the grant \emph{``Finite mixture and latent variable models for causal inference and analysis of socio-economic data''} (FIRB - Futuro in ricerca) funded by the Italian Government (RBFR12SHVV).

\section*{Appendix}
\label{sec:weights}
The sampling scheme adopted by TIMSS and PIRLS is a stratified two-stage cluster sample design, with schools as primary units and classes as secondary units \citep{martin:12}.
Schools are stratified in a country-specific way. In Italy, explicit strata are defined by combining the geographical area (7 strata) and the classification of schools into `Grade 4 only' and `Grade 4 \& 8' (2 strata).  Implicit strata are also considered, they are defined by sorting the schools  within each explicit stratum according to the  type of school (private or public, 2 strata) and  province (104 strata).
The schools are sampled with a random-start fixed-interval systematic scheme, with each school selected with probability proportional to its size (PPS).
In the second stage, for each school one or two classes are sampled with equal probability (after grouping together classes with size below a given threshold). All the students of the sampled classes take part in the survey.

In a regression model, weights are needed to obtain unbiased estimates when the sampling is informative, namely the inclusion probabilities are related with the model errors, which is an assumption not directly verifiable. Unfortunately, sample weights inflate the standard errors of the estimators, thus the trade-off between bias and variance should be evaluated case by case. The role of weights is usually assessed by comparing model estimates with and without weights, see among others \cite{Rabe:06}. 
At each hierarchical level, the weight is defined as the product of the \textit{sampling weight} (i.e. the reciprocal of the conditional sampling probability) and the \textit{adjustment weight}, which accounts for non-participation of sampled units. The overall student weight is obtained by multiplying the weights across the three hierarchical levels (student $i$, class $j$, school $k$): $w_{ijk}=w_{i \mid jk} w_{j \mid k} w_{k}$. In order to perform weighted estimation in multilevel models, the weights must refer to the relevant hierarchical levels \citep{Rabe:06}. 

We checked the effect of including sample weights by fitting univariate multilevel models with and without weights for each score separately.
Therefore, in our two-level model with students nested into classes, we insert both the conditional student weight $w_{i \mid jk}$ and the unconditional class weight $w_{jk}= w_{j \mid k} w_{k}$.
Considering the score on Reading, the use of sample weights slightly affects the estimates of the regression coefficients: the signs are unchanged and the average absolute relative variation amounts to 8.7\%. However, as expected, the use of sample weights inflates the standard errors for most regression coefficients, with an average absolute relative variation of 7.2\%.
The ICC reduces from 0.15 to 0.13. Analogous patterns are observed for Math and Science.
Overall, we conclude that the use of sample weights do not change the substantial findings of the analysis.
Moreover, combining weighted estimation and multiple imputation would yield large standard errors. 
This is the reason why in Section  \ref{sec:findings} we choose to report  the results of the model with the unweighted estimation.

\bibliography{biblio2}
\bibliographystyle{apalike}

\end{document}